\documentclass[prd,superscriptaddress,nofootinbib]{revtex4}
\usepackage{amssymb,amsmath,graphicx,amscd,epsfig}
\usepackage[colorlinks=true, pdfstartview=FitV, linkcolor=blue, citecolor=red, urlcolor=magenta, breaklinks=true]{hyperref}

\newcommand{\be}{\begin{equation}}
\newcommand{\ee}{\end{equation}}
\newcommand{\beqq}{\setlength\arraycolsep{2pt}\begin{eqnarray}}
\newcommand{\eeqq}{\vspace{0cm} \end{eqnarray}}
\newcommand{\bea}{\begin{eqnarray}}
\newcommand{\eea}{\end{eqnarray}}

\begin{document}

\title{LIV effects on
the quantum stochastic motion in an acoustic FRW-geometry}

\author{M. A. Anacleto} \email{anacleto@df.ufcg.edu.br}
\affiliation{Departamento de F\'{\i}sica, Universidade Federal de Campina Grande 
Caixa Postal 10071, 58429-900 Campina Grande, Para\'iba, Brazil}

\author{C. H. G. Bessa } \email{chgbessa@yahoo.com}
\affiliation{Departamento de F\'{\i}sica, Universidade Federal de Campina Grande 
Caixa Postal 10071, 58429-900 Campina Grande, Para\'iba, Brazil}

\author{F. A. Brito}\email{fabrito@df.ufcg.edu.br}
\affiliation{Departamento de F\'{\i}sica, Universidade Federal de Campina Grande
Caixa Postal 10071, 58429-900 Campina Grande, Para\'iba, Brazil}
\affiliation{Departamento de F\'isica, Universidade Federal da Para\'iba, 
Caixa Postal 5008, 58051-970 Jo\~ao Pessoa, Para\'iba, Brazil}

\author{A. E. Mateus}  \email{andersoneva20@gmail.com}
\affiliation{Departamento de F\'{\i}sica, Universidade Federal de Campina Grande 
Caixa Postal 10071, 58429-900 Campina Grande, Para\'iba, Brazil}

\author{E. Passos}\email{passos@df.ufcg.edu.br}
\affiliation{Departamento de F\'{\i}sica, Universidade Federal de Campina Grande
Caixa Postal 10071, 58429-900 Campina Grande, Para\'iba, Brazil}

\author{J. R. L. Santos}\email{joaorafael@df.ufcg.edu.br}
\affiliation{Departamento de F\'{\i}sica, Universidade Federal de Campina Grande
Caixa Postal 10071, 58429-900 Campina Grande, Para\'iba, Brazil}

\begin{abstract}

It is well known in the literature that vacuum fluctuations can induce a random motion of particles which is sometimes called quantum Brownian motion or quantum stochastic motion. In this paper, we consider Lorentz Invariance Violation (LIV) in an acoustic spatially flat Friedman-Robertson-Walker (FRW) geometry.  In particular, we are looking for the LIV effects in the stochastic motion of scalar and massive test particles. This motion is induced by a massless quantized scalar field on this geometry, which in turn is derived from an Abelian Higgs model with LIV. Deviations in the velocity dispersion of the particles proportional to the LIV parameter are found.

\end{abstract}

\maketitle

\pretolerance10000

\section{Introduction}

It is well accepted today that in a complete theory of quantum gravity a fundamental concept like the Lorentz invariance (LI), which arises naturally from Lorentz transformations in special relativity, should be violated in some regime. In general, it is quite believed that this violation, also known as the Lorentz invariance violation (LIV), should appear in the energy range close to the Planck scale. It was Kostelecky and collaborators \cite{ks89a} who proposed one of the first effective theories where the LIV was permitted. They developed a model that extends the standard model into a new one where the LIV could appear. In this extended model, all sectors are corrected by small coefficients or parameters which can be adjusted according to the requirements of the theory under consideration. The search for phenomenological models where the LIV could be detectable was developed by many authors \cite{ks89a,ks89b,ks91,kl01,ck98,bk04,k03,bbm05,abp10,abp11,Anacleto:2019rfn}. Although no violation has appeared so far, such studies have proved to be very useful once they enable us to stipulate the values of the LIV parameters. In this context, we mention works related to quantum gravity effects at low energy scales. Among many possibilities we call attention to studies related directly to  string theory \cite{ks89a,ks89b,ks91}, loop quantum gravity \cite{atu00}, Horava-Lifshitz theory \cite{h09},  noncommutativity \cite{chklo01,bbq08,n09,bl16}, and also models related to modified gravity \cite{lsma20} and special relativity \cite{cg06,msca14,ms02}. All of these approaches are attempts to search for LIV effects that could be in principle testable with future technologies. 

In this framework, the main purpose of this paper is to show a different model where LIV effects could appear.  The model consists of an acoustic spatially flat Friedmann-Robertson-Walker (FRW) geometry derived from an Abelian Higgs model with LIV in the scalar sector. In this scenario, we study the motion of a scalar test particle under the influence of a massless quantized and vacuum fluctuating scalar field. It is well known in the literature that vacuum fluctuations can induce a random motion of particles which is sometimes called quantum Brownian motion\footnote{To avoid confusion with other models which use the same nomenclature in but are related to non-equilibrium quantum thermodynamic models \cite{ehm09,bcgaa18,dc19}, we decide to call this effect here as quantum stochastic motion.}, or quantum stochastic motion. So, the type of motion we will deal with in the present paper is characterized by the calculation of the velocity dispersion of the particles ($\langle \Delta v^2\rangle$) where a Langevin type equation is considered and the particles dispersion velocity is directly associated with the Hadamard function of the quantum fields. Thus, in a LI perspective, this stochastic motion in Minkowski spacetime is exhibited only when some physical boundary or non-trivial topology is present \cite{gs99,jr92,YuFord2004,YuChen2004,Seriu2009,Seriu2008,Lorenci2019,clrr19,pf11,lrs16,clrrs18,cy04,zy05,br20,mm19,DaShinLee2005,DaShinLee2008,DaShinLee2014}, but for a time-dependent expanding background, as a FRW type, it could appear without any boundary intervention and with trivial topology \cite{Bessa2017,Bessa2009,abbfp20}. Particularly, for the scalar particles, which are the kind of particles we are interested in, this stochastic motion shows up only when they are submitted to an external classical non-fluctuating force ($f_{ext}$). In this scenario, these particles are usually named as `bound particles' once this $f_{ext}$ can prevent them to feel locally the expansion. In a different and more realistic scenario, this external force could vanish and the particles could follow their geodesics. In this case, the particles are named `free particles' and their stochastic motion is null once their velocity dispersion is zero ($\langle \Delta v^2\rangle = 0$), {\it {i.e.}}, the free particles present a dispersion identical to the one found in the  Minkowski case without boundaries and with trivial topology. In this context, the main result of this paper is to show that when we take into account the LIV model, the quantum stochastic motion appears for both types of particles discussed above, {{\it i.e.}} $\langle \Delta v^2\rangle_{LIV} \neq 0$, for bound and free particles. A similar result was found in Ref. \cite{abbfp20} for a (3+1)-dimensional non-commutative Abelian Higgs model.

So this article is organized as follows: in Sect. \ref{Sect:review} we review some basic results found in Ref. \cite{Bessa2017} where the study of the stochastic motion for scalar particles was developed in an expanding Bose-Einstein condensate (BEC) that mimics the FRW geometry in the LI context. Then in Sect. \ref{Sec:LIV} we show, following the steps described in Ref. \cite{abp10}, how to find a similar acoustic FRW geometry from an Abelian Higgs model with LIV terms. This new acoustic geometry will be used in Sect. \ref{sect:corrections} to reanalyze the problem discussed in Sect. \ref{Sect:review} but now in a LIV perspective. In this case, small corrections in the particle's velocity dispersion proportional to the LIV parameters are found when $f_{ext} \neq 0$ and $f_{ext} = 0$. In Sect. \ref{Sect:conclu} we discuss the main results and give the conclusions. In this paper we use units where $\hbar = c = 1$.

			\section{Quantum stochastic motion in an acoustic FRW geometry: Short review}\label{Sect:review}

				In this section, we review the main results of Ref. \cite{Bessa2017} where the stochastic motion of a scalar point particle that was under the influence of a quantized vacuum fluctuating and massless scalar field in a FRW geometry was considered. This study was carried out in a LI framework using a Bose-Einstein condensate (BEC). In this context, a well-known result is the possibility to mimic some aspects of a FRW universe using superfluids with remarkable experimental results \cite{ekjsc18}. Thus, in Refs. \cite{Bessa2017,Visser:1997ux,Barcelo2011,bffr08,jwvg07,pfl10}, a linearization in the basic equation of motion of an inviscid and irrotational BEC was applied. This linearization was done in the fluid constituents such as the field $\phi = \phi_0 + \phi_1$, the phase $S = S_0 + S_1$,  fluid density $\rho = \rho_0 + \rho_1$, and with a vanished velocity flow $\vec{v}=\hbar\nabla\phi_0/m = 0$. Following this procedure, {we can find an equation of motion similar to a Klein-Gordon equation in curved spacetime
\begin{equation}
\Box \phi_1 = 0,
\end{equation}				
where $\Box = \frac{1}{\sqrt{-g}}\partial_\mu(\sqrt{-g}g^{\mu\nu}\partial_\nu)$ with an effective spatially flat FRW metric:	}
\be
\label{Metrica0}
ds_{eff}^2=-c_0^2dt^2+ a_{eff}^2(t)(dx^2+dy^2+dz^2) ,
\ee		
where the dimensionless effective scale factor is $a_{eff}(t) = (c_0/c_s(t))^2$, with $c_s(t)=4\pi\hbar\rho_0 L(t)/m$ been the time dependent sound speed, $c_0 = c_s(t=t_0)$, the sound speed in certain initial time, and $L(t)$ a time dependent scattering length. Note that, in order to have an analog of expanding universe, $a_{eff}$ must increase with time when $c_{s}(t)$ decreases. As usual, a conformal transformation of the type, $dt = ad\eta$, is possible to be made\footnote{For briefness, we write from now on $ds_{eff} = ds$ and $a_{eff} = a$.}. Thus we obtain a conformal FRW analog geometry represented by: 			
\be
\label{Metrica1}
ds^2=a^2(\eta)\left(-c_0^2d\eta^2+ dx^2+dy^2+dz^2\right) ,
\ee
where $\eta$ is the conformal time.

In this model, the atoms that compound the fluid are treated as scalar point particles that interact with a quantized massless scalar field, which represents the acoustic perturbations driven by the phonons in this effective time-dependent geometry. Thus, as was pointed out in Ref. \cite{Bessa2017}, these particles fell the fluid expansion according to the effective metric (\ref{Metrica0}) or (\ref{Metrica1}). In this context, the quantized field could induce metric fluctuations. However, the contributions that come from these fluctuations are usually secondary to the motion of the point particles considered in these types of models \cite{abbfp20}

The  equation of motion for a single particle in certain $i$-direction is given by
\begin{equation}\label{eqmotion0}
m\frac{Du^i}{d\tau} = {f'}^{i}  ,
\end{equation}
where $i \in \{x,y,z\}$, $D/d\tau$ is a covariant derivative, $u^i$ is the coordinate $3$-velocity, and $m$ is the particle mass. From equation above, the left hand side is 
\begin{equation}
\label{2LN3}
m\dfrac{du^i}{dt}+2m\dfrac{\dot{a}}{a}u^i= {f'}^{i},
\end{equation}
where the proper time $\tau \approx t$, for non-relativistic motion, and Eq. (\ref{Metrica0}) was used to calculate the Cristofell symbols $\Gamma^i_{\alpha\beta} = \dot{a}/a$. In the right hand side of Eq. (\ref{2LN3}), the force term ${f'}^{i}$ can be split into two parts, as follows: ${f'}^{i} = f^{i}+ f_{ext}^i $. Here, $f^i$ is a term which is directly related with a  scalar field { and it can be derived from the Klein-Gordon equation as \cite{ppv11}} 
\begin{equation}\label{eqmotion1}
f^i = qg^{i\nu}\nabla_\nu\phi_1,
\end{equation}
where is valid for a flat or conformally flat spacetime \cite{Bessa2017,ppv11},  the covariant derivative is the partial derivative ($\nabla_\mu\phi = \partial_\mu\phi$), and $q$ is a scalar charge which gives the strength of the interaction between particles and the field. The other term, $f_{ext}^i$, is a classical external non-fluctuating force. In fact, $f_{ext}^i$ can be considered equal to zero, without loss of generality. In this case, the particles are called `free particles', in such way that they follow their own geodesic. However, in some situations $f_{ext}$ can be different from zero, or more conveniently $f_{ext} \approx \dot{a}/a$, in such a way that the particles do not feel locally the expansion. In this case they are named `bound particles'. This nomenclature was adopted in Refs. \cite{Bessa2009, Bessa2017, abbfp20} and we will follow it in the present paper.

\subsection{Bound particles}


Let us start reviewing the bound particles case. For this purpose consider that the scalar particles are under the influence of an external force $f_{ext} = 2m\frac{\dot{a}}{a}u^i$, in such way that Eq. (\ref{2LN3}) reduces to
\begin{equation}
\frac{du^i}{dt} = \frac{f^i}{m}.
\end{equation}

After integrating the equation above and considering a zero initial velocity, $u^i(t_0) = 0$, we can evaluate the dispersion for the coordinate velocity of the scalar particle. Thereafter, we evaluate the dispersion for the physical $3$-velocity, namely, proper velocity ($v^i$), in this expanding geometry. The relation between $u^i$ and $v^i$ is well known and it is associated with the coordinate length ($r$)  and proper length ($l$), that is, $l = a_fr$, where the sub-indexes indicates that the scale factor was evaluated in a final time ($t_f$), {\it{i.e.}}, when the expansion ceases. Thus, $v^i = a_fu^i$ and taking into account Eq. (\ref{eqmotion1}), we obtain 
\begin{equation}
\label{cor4}
\langle (\Delta v^{i})^2\rangle=\dfrac{q^2 a_f^2}{m^2}\left[\partial_{i_1}\partial_{i_2}\int_{\eta_i}^{\eta_f}\int_{\eta_i}^{\eta_f} d\eta_1 d\eta_2 a^{-2}(\eta_1) a^{-2}(\eta_2) \langle \phi_1(\eta_1, r_1)\phi_1(\eta_2, r_2)\rangle_{M}\right]_{r_1\rightarrow r_2}.
\end{equation}
This is the proper velocity dispersion\footnote{ The velocity dispersion was defined as usual: $\langle (\Delta v^{i})^2\rangle = \langle v^i(r_1, t_1)v^i(r_2, t_2)\rangle - \langle v^i(r_1, t_1)\rangle\langle v^i(r_2, t_2)\rangle$} where a change between coordinate time $t$ and conformal time $\eta$ was made and the coincidence limit in the spatial coordinates ($r_1 \rightarrow r_2$) was taken, after the spatial partial derivatives were applied. Note that the scalar field $\phi$ obeys the relation $\langle \phi(t_1, r_1)\phi(t_2, r_2)\rangle \neq 0, \langle \phi(t, r)\rangle = 0$. The sub-indexes $M$, indicates that the correlation function for the massless scalar field was evaluated in a $4$-dimensional Minkowski spacetime and by the conformal metric (\ref{Metrica1}) it is related to the $FRW$ correlation function by a conformal factor  $\Omega$ \cite{Birrell1984}, that comes from the relation $g_{\mu\nu} = \Omega^2\eta_{\mu\nu}$. From Eq. (\ref{Metrica1}), $\Omega = {a(\eta)}$, thus, from Ref. \cite{Birrell1984}, we obatin,
\begin{equation}\label{correlation}
\langle \phi_1(\eta_1, r_1)\phi_1(\eta_2, r_2)\rangle_{FRW} = \Omega^{-1}(\eta_1, r_1)\Omega^{-1}(\eta_2,r_2)\langle \phi(\eta_1, r_1)\phi(\eta_2, r_2)\rangle_{M} =a^{-1}(\eta_1)a^{-1}(\eta_2)\langle \phi_1(\eta_1, r_1)\phi_1(\eta_2, r_2)\rangle_{M}. 
\end{equation}
Where $\langle \phi_1(\eta_1, r_1)\phi_1(\eta_2, r_2)\rangle_{M}$ is the Hadamard function in Minkowski spacetime, which is given by:
\begin{equation}\label{hadamard}
\langle \phi_1(\eta_1, r_1)\phi_1(\eta_2, r_2)\rangle_{M} = \frac{1}{4\pi^2}\frac{1}{[-c_0^2(\eta_1-\eta_2)^2 + r^2]},
\end{equation}
with $r^2 = \Delta x^2 + \Delta y^2 + \Delta z^2$.

To integrate Eq. (\ref{cor4}) we must choose an appropriate effective scale factor $a(\eta)$. In many papers \cite{Bessa2017,abbfp20,jwvg07,pfl10} the authors consider an  asymptotically bound dimensionless scale factor to model the effective $FRW$ geometry\footnote{Recently, a model that reproduces in the laboratory some features of an inflationary universe was performed \cite{ekjsc18} with a scale factor similar to Eq. (\ref{tanh}) }, whose form is 
\begin{equation}\label{tanh}
a^2(\eta) = a_0^2 + a_1^2\tanh\left(\frac{\eta}{\eta_0}\right),
\end{equation}        
where $a^2(\eta)$ is flat in the asymptotic regions $(\eta \rightarrow \pm\infty)$ and $a_0$, $a_1$ are dimensionless constants, with the former producing a displacement of $a(\eta)$ avoiding the divergence when $\eta = 0$ and the latter modifying  the space in the limits $\eta \rightarrow \pm \infty$. The $\eta_0$ is a constant parameter with dimension of time that smooths the transition between the asymptotic limits.  Thus, we can evaluate Eq. (\ref{cor4}) using this scale factor. Proceeding this way, we take Eqs. (\ref{hadamard}) and (\ref{tanh}) in Eq. (\ref{cor4}), and to integrate it we followed the same steps present in Appendix B of Ref. \cite{Bessa2009}. Then, the proper velocity dispersion is
\begin{equation}
\label{caso ligado}
\langle(\Delta v^i)^2\rangle = \dfrac{2 q^2 B}{\pi^4 m^2 c_0^4 \eta_0^2}\left(\zeta(3)- \dfrac{\pi^4}{90}\right) ,
\end{equation}
where 
\begin{equation}\label{eq:B}
B = \frac{c_0}{4c_{sf}}\left(1 - \frac{c_{sf}}{c_0}\right)^2,
\end{equation}
is a dimensionless constant defined in terms of the initial ($c_0$) and final ($c_{sf}$) sound speed, besides $\zeta(n)$ is the zeta function. Note that, when $c_{sf} = c_0$, $B=0$ a null dispersion is found. This is the expected result of a Minkowski spacetime with trivial topology \cite{YuFord2004,YuChen2004,Bessa2009}.

In the Sect. \ref{sect:corrections} we will re-analyse this problem and will find extra terms (corrections) to the velocity dispersion coming from an expanding  LIV model similar to  Eq. (\ref{Metrica1}).

\subsection{Free particles}

The second case studied in Ref. \cite{Bessa2017} was the quantum stochastic motion of the free particles. As previously presented, this case is characterized by the fact that $f_{ext} = 0$. So, Eq. (\ref{2LN3}) can be written as
\begin{equation}
\frac{1}{a^2}\frac{d}{dt}(a^2u^i) = \frac{f^i}{m},
\end{equation}
where $f^{i}$ is given by Eq. (\ref{eqmotion1}).

To integrate this equation and to find the proper velocity dispersion ($(\Delta v^i)^2$) we follow the same procedure discussed in the bound particles section. Thus, the velocity dispersion is 
\begin{equation}
\label{cor12}
\langle (\Delta v^{i})^2 \rangle= \dfrac{q^2}{m^2 a_f^2}\partial_{i1}\partial_{i2}\left[\int_0^{\eta_f}\int_0^{\eta_f} d\eta_1 d\eta_2 \langle \phi_1(\eta_1, r_1) \phi_1(\eta_2, r_2) \rangle_{M}\right]_{r_1\rightarrow r_2} .
\end{equation}
Note that the scale factor $a$ does not appears in the integrand above. Using Eq. (\ref{hadamard}) and taking the coincidence limit after the integration, we find a divergent result. To avoid this divergence, a renormalization is required. It consists in subtract the Minkowski contribution in the Hadamard function. Thus, the integrand will be null yielding to the trivial result 
\begin{equation}\label{lorentzfree}
\langle(\Delta v^i)^2 \rangle = 0 .
\end{equation}
In other words, there is no stochastic motion for free scalar particles when the LI is preserved.

In the Sect. \ref{sect:corrections} we will take into account the LIV model in the expanding geometry. We will find a non-null velocity dispersion. This indicates that the model proposed in this paper could be a source to look for LIV effects.

			
			\section{Analog model to the FRW geometry in a LIV perspective}\label{Sec:LIV}

In the previous section, we have reviewed the formalism related to the quantum stochastic motion of scalar particles. This motion is induced by the quantum vacuum fluctuations of a massless scalar field in an expanding time-dependent background. It was discussed that this background can mimic some aspects of a FRW geometry when a linearization in the parameters of the basic equation of a BEC is applied. Since, as far as we know, the Lorentz invariance is preserved in a BEC,  our goal in this section is to show that a similar time-dependent expanding background can be accomplished when we take into account a model where LIV is permitted. For this purpose, we consider an extension in the Abelian Higgs model with a term that violates the Lorentz invariance in the scalar sector. We will follow the procedure present in Section 2 of Ref. \cite{abp10}. Thus, consider a Lagrangian of the type,
\be
\label{lagrangiana}
\mathcal{L} = -\dfrac{1}{4}F_{\mu\nu}F^{\mu\nu}+|D_{\mu}\phi|^2 + M^2|\phi|^2 - b|\phi|^4 + k^{\mu\nu}D_{\mu}\phi^{*}D_{\nu}\phi\,,
\ee
with
\be
k_{\mu\nu}= \begin{pmatrix}
\beta & \alpha & \alpha & \alpha\\
\alpha & \beta & \alpha & \alpha\\
\alpha & \alpha & \beta & \alpha\\
\alpha & \alpha & \alpha & \beta\\
\end{pmatrix}. 
\ee
{The tensor $k_{\mu\nu}$  introduces the Lorentz symmetry breaking terms, represented by the real parameters $\alpha$ and $\beta$, which were coupled to the scalar field.} In this Lagrangian, gravity is not present and the term $F^{\mu\nu}$ is the Maxwell tensor, $D_{\mu} = \partial_\mu - ieA_\mu$ with $e$ and $b$ being interaction terms, and $A_{\mu}$ is the $4$-potential. 

In order to simplify the model, we will only consider the diagonal terms in $k_{\mu\nu}$ matrix, {\it{i.e.}} $\alpha = 0, \beta \neq 0$. A decomposition of the type: $\phi(x',t')=\sqrt{\rho(x',t')}exp(iS(x',t'))$, {which is known as Madelung representation, can be done with the prime representing the coordinates in the flat spacetime Lagrangian. Notice that the fluid density is now defined as $\rho = |\phi|^2$. This representation gives us a fluid description and after substituting $\phi$ in Eq. (\ref{lagrangiana}) we obtain the corresponding hydrodynamic equations of motion:} 
\be
\label{continuidade}
-\partial_{t'} \left[\tilde{\beta}_+\rho (\dot{S} -eA_{t'})\right] +\partial_{i'}\left[\tilde{\beta}_-\rho (\partial^{i'}S -eA^{i'})\right]=0
\ee
and
\be
\label{fluido}
\dfrac{(\tilde{\beta}_+\partial_{t'}^2 - \tilde{\beta}_-\partial_{i'}^2)\sqrt{\rho}}{\sqrt{\rho}} +\tilde{\beta}_+(\dot{S} -eA_{t'})^2 - \tilde{\beta}_-(\partial_{i'}S -eA_{i'})^2 +M^2 -2b\rho=0, 
\ee
where $\tilde{\beta}_{\pm}\equiv 1 \pm \beta$ and $\dot{S}=\frac{\partial S}{\partial t'}$.

{Let us now consider a perturbation around the density given by $\rho=\rho_0+\rho_1$. We also can change the phase for $S=S_0 + S_1$, and consequently we have $\phi=\phi_0+\phi_1$, where $\rho_1 \ll \rho_0$, $S_1 \ll S_0$, and $\phi_1 \ll \phi_0$. So for simplicity, $M = 0$, equations above become}
\begin{eqnarray}\label{eqnew14}
&\partial_{t'} &\left[\left(-\dfrac{b\rho_0}{2\tilde{\beta}_-}+ \dfrac{\mathcal{D}_2\rho_0}{2\tilde{\beta}_-}-\dfrac{\tilde{\beta}_+}{\tilde{\beta}_-}\omega_0^2\right) \dot{S}_1 - \omega_0\vec{v}_0 \cdot\nabla S_1\right]\nonumber\\
&+& \nabla \cdot \left[-\omega_0\vec{v}_0 \dot{S}_1 + \left(\dfrac{b\rho_0}{2\tilde{\beta}_+}-\dfrac{\mathcal{D}_2\rho_0}{2 \tilde{\beta}_+}\right)\nabla S_1 - \dfrac{\tilde{\beta}_-}{\tilde{\beta}_+}\vec{v}_0 \cdot \nabla S_1 \vec{v}_0\right]=0.
\end{eqnarray}
{These fluctuations
of the fluid are similar to the BEC case studied in Sect. \ref{Sect:review} and in other analogue gravity models \cite{Barcelo2011}. Furthermore, we obtain a phonon description when these fluctuations (or equivalently $\phi_1$) are quantized, as we shall see shortly.} Note that we have defined $\mathcal{D}_2 = \tilde{\beta}_+ D_{t2} + \tilde{\beta}_- D_{i2}$, $\omega_0=-\dot{S}_0 + eA_t$ and $\vec{v_0}= \nabla S_0 + e\vec{A}$ is the local velocity field.  The term $\mathcal{D}_2$ is very small and following Ref. \cite{abp10} it can be dropped out. 

Following the usual procedure, we can see that Eq. (\ref{eqnew14}) mimics a Klein-Gordon equation in curved spacetime with the metric

{
\begin{equation}
d\bar{s}^2 = \frac{b\rho_0(\tilde{\beta}_{-})^{1/2}}{2c_s}\left[-\frac{c_s^2}{\tilde{\beta}_{+}} d{t'}^2 + \frac{\tilde{\beta}_+}{\tilde{\beta}_-}\left(d{x'}^2 +d{y'}^2 + d{z'}^2\right)\right].
\end{equation}
{Here we have assummed a non-relativistic limit where the time dependent sound speed in the fluid is $c_s^2 = \frac{b\rho_0}{2w_0^2} \ll 1$, and the background flow is null ($\vec{v} = 0$).} 
}

Absorbing the constants above into the coordinates $t', x', y', z'$ and in $d\bar{s}$ we find the usual flat metric evaluated in a Lorentz invariant (LI) context
\begin{equation}
d\bar{s}^2 = -c_s^2d{t'}^2 + d{x'}^2 +d{y'}^2 + d{z'}^2.
\end{equation}
This is clearly different from the metric obtained in Ref. \cite{abp10}, in which the flux velocity does not vanish, and an acoustic Kerr-like black hole metric with the LIV corrections is found. Despite of the LI metric above,  we will see that the $\beta$ factor present in Eq. (\ref{lagrangiana}) influences the equation of motion of the particles anyway. 
%
%
Finally, to write this metric in FRW form (\ref{Metrica1}), we simply follow the discussion presented in Sect.~\ref{Sect:review}. {In this case we have a Klein-Gordon equation in curved spacetime with}
{
\begin{equation}\label{KGLIV}
\frac{1}{\sqrt{-g}}\partial_\mu(\sqrt{-g}\tilde{g}^{\mu\nu}\partial_\nu\phi_1) = 0,
\end{equation}
where the effective metric tensor is now $\tilde{g}^{\mu\nu} = g^{\mu\nu} + k^{\mu\nu}$.}

In the next section we will use this metric to study the quantum stochastic motion of the scalar particle described in Sect. \ref{Sect:review}. 

				\section{Quantum stochastic motion in a FRW acoustic geometry: LIV effects}\label{sect:corrections}
				
{In this section, we consider Eq. (\ref{KGLIV}), and  follow the same methodology described in Sect. \ref{Sect:review} to find a new $f^i$ term. Now, it will take into account the LIV corrections in the following way:}
\begin{equation}
\label{eq5}
f^i=q({g}^{i\mu}+k^{i\mu})\partial_\mu\phi_1.
\end{equation} 
Note that when $k^{\mu\nu} = 0$, Eq. (\ref{eqmotion1}) is recovered, which is valid for a flat or conformally flat spacetime where LI is preserved. Since we use metric (\ref{Metrica0}), the left hand side of Eq. ($\ref{2LN3}$) does not change. So we have now the following equation of motion 
\begin{equation}\label{eq:motionLIV}
m\dfrac{du^i}{dt}+2m\dfrac{\dot{a}}{a}u^i= q[a^{-2}(t)+\beta]\partial_i\phi_1 + f^i_{ext},
\end{equation}
{where $\dot{a} = da/dt$. This is the equation of motion of the scalar particle taking into account the LIV effects where $\beta$ comes from Eq. (\ref{KGLIV}).}

{In this context, in the next two sub-sections, we will consider the acoustic metric evaluated in the previous section to study the stochastic motion of point-like massive scalar particles in this expanding background that mimics some features of a spatially flat FRW geometry. Thus, to evaluate the dispersion velocity of the particles with mass $m$ of the compounds (treated as scalar point particles) of the fluid with density $\rho$, we will consider again the bound and free particles defined in Sect. \ref{Sect:review}. 
}


\subsection{Bound Particles}

Let us start this section with the bound particles case, as discussed in Sect. \ref{Sect:review}, a single particle is under the effect of a classical non-fluctuating force given by: $f^i_{ext} = 2mu^i\dot{a}/a$. Thus the equation of motion for the particles reduces to
\begin{equation}
m\dfrac{du^i}{dt}= q[a^{-2}(t)+\beta]\partial_i\phi_1,
\end{equation} 
where only the spatial derivatives are considered once the metric $g^{\mu\nu}$ and the $k^{\mu\nu}$ matrix are diagonal. Thus,  the corresponding velocity-velocity correlation function using the LIV effective metric is
\begin{eqnarray}\label{eq:uu}
\langle u^i(t_1, r_1)u^i(t_2, r_2)\rangle_{LIV} = \frac{q^2}{m^2}\int\int dt_1dt_2\left\{\left[a^{-2}(t_1) + \beta\right]\left[a^{-2}(t_2) + \beta\right]\right\} \\ \nonumber \times \partial_{i_1}\partial_{i_2} \langle \phi_1(t_1, r_1)\phi_1(t_2, r_2)\rangle_{FRW}
\end{eqnarray}

Following the procedure presented in Sect. \ref{Sect:review}, we first make a conformal transformation involving the coordinate time $t$ and the conformal time $\eta$ ($dt=ad\eta$) and make use of Eqs. (\ref{correlation}) and (\ref{Metrica1}) to obtain the two-point function in the FRW spacetime in terms of the Hadamard function in Minkowski spacetime
\begin{equation}\label{Wcorrelation}
\langle \phi_1(\eta_1, r_1)\phi_1(\eta_2, r_2)\rangle_{FRW} = a^{-1}(\eta_1)a^{-1}(\eta_2)\langle \phi_1(\eta_1, r_1)\phi_1(\eta_2, r_2)\rangle_{M}, 
\end{equation}

Using Eq. (\ref{Wcorrelation}) in Eq. (\ref{eq:uu}) together with the change, $dt=ad\eta$, we obtain,
\begin{eqnarray}
\label{eq10}
\langle \langle u^i(t_1, r_1)u^i(t_2, r_2)\rangle\rangle_{LIV} &=& \dfrac{q^2  }{m^2}\int d\eta_2 a^{-2}(\eta_2)\int d\eta_1 a^{-2}(\eta_1)\partial_{i_1}\partial_{i_2}\langle\phi_1(\eta_1, r_1)\phi_1(\eta_2, r_2)\rangle_{M}+ \nonumber\\
&+& \dfrac{q^2\beta }{m^2}\int d\eta_2\int d\eta_1 a^{-2}(\eta_1)\partial_{i_1}\partial_{i_2}\langle\phi_1(\eta_1, r_1)\phi_1(\eta_2, r_2)\rangle_{M}+ \nonumber\\
&+& \dfrac{q^2\beta }{m^2}\int d\eta_1\int d\eta_2 a^{-2}(\eta_2)\partial_{i_1}\partial_{i_2}\langle\phi_1(\eta_1, r_1)\phi_1(\eta_2, r_2)\rangle_{M} + \nonumber\\
&+& \dfrac{q^2\beta^2 }{m^2}\int d\eta_1\int d\eta_2 \partial_{i_1}\partial_{i_2}\langle\phi_1(\eta_1, r_1)\phi_1(\eta_2, r_2)\rangle_{M}.
\end{eqnarray}

Note that the scale factor does not appears in the last term of Eq. (\ref{eq10}). Thus, the integration diverges in the coincidence limit. This implies that a renormalization is needed. So the Minkowski contribution must be subtracted from Hadamard function presented in the last integral, giving a null contribution. Thus, the remained finite integrals are 
\begin{eqnarray}
\label{eq11}
\langle v^i(t_1, r_1)v^i(t_2, r_2)\rangle_{LIV} &=& \dfrac{q^2 a_f^2}{m^2}\int_{-\infty}^\infty d\eta_2 a^{-2}(\eta_2)\int_{-\infty}^\infty d\eta_1 a^{-2}(\eta_1)\partial_{1i}\partial_{2i}\langle\phi_1(\eta_1, r_1)\phi_1(\eta_2, r_2)\rangle_{M}+ \nonumber\\
&+& \dfrac{q^2 a_f^2\beta}{m^2}\int_{0}^{\eta_f} d\eta_2\int_{-\infty}^\infty d\eta_1 a^{-2}(\eta_1)\partial_{1i}\partial_{2i}\langle\phi_1(\eta_1, r_1)\phi_1(\eta_2, r_2)\rangle_{M}+ \nonumber\\
&+& \dfrac{q^2 a_f^2\beta}{m^2}\int_{0}^{\eta_f} d\eta_1\int_{-\infty}^\infty d\eta_2 a^{-2}(\eta_2)\partial_{1i}\partial_{2i}\langle\phi_1(\eta_1, r_1)\phi_1(\eta_2, r_2)\rangle_{M}, 
\end{eqnarray}
where we considered the proper velocity ($v^i = a_fu^i$).

Particularizing $i$ to the $x$ direction, the results in the $y$ and $z$ direction are the same, and taken the partial derivatives with respect to $x_1$ and $x_2$, we obtain,
 \begin{eqnarray}
\label{eq11a}
\langle v^i(t_1, r_1)v^i(t_2, r_2)\rangle_{LIV} &=& \dfrac{q^2a_f^2}{2\pi^2m^2}\int_{-\infty}^\infty d\eta_2 a^{-2}(\eta_2)\int_{-\infty}^\infty d\eta_1 a^{-2}(\eta_1)\left\{f_2(\eta,r)+4\Delta x^2f_3(\eta, r)\right\}+ \nonumber\\
&+& \dfrac{q^2 a_f^2\beta}{m^2}\int_{0}^{\eta_f} d\eta_2\int_{-\infty}^\infty d\eta_1 a^{-2}(\eta_1)\left\{f_2(\eta,r)+4\Delta x^2f_3(\eta, r)\right\}+ \nonumber\\
&+& \dfrac{q^2 a_f^2\beta}{m^2}\int_{0}^{\eta_f} d\eta_1\int_{-\infty}^\infty d\eta_2 a^{-2}(\eta_2)\left\{f_2(\eta,r)+4\Delta x^2f_3(\eta, r)\right\}, 
\end{eqnarray}
where $f_n$ was defined as,
\begin{equation}\label{eq:fn}
f_n(\eta, r) = \frac{1}{[c_0^2(\eta_1 - \eta_2)^2 - r^2]^n}.
\end{equation}

Now we use the scale factor given by Eq. (\ref{tanh}) and to evaluate the integral above we follow again a contour integration presented in appendix B of Ref. \cite{Bessa2009}. The result of such integral is:
\begin{eqnarray}\label{eq_LIV45}
\langle(\Delta v^x)^2\rangle_{LIV} = \frac{2q^2B}{\pi^4m^2c_0^4\eta_0^2}\left(\zeta(3) - \frac{\pi^4}{90}\right) + \frac{2q^2\beta}{3c_0^4\pi m^2\eta_0^2}\left(\frac{c_0}{c_{sf}}B\right)^{1/2}F_k(\eta),
\end{eqnarray}
where the coincidence limit was taken in the expression above and
\begin{equation}
\label{eq24}
F_k(\eta) = \dfrac{1}{2 \pi ^3}\left[\psi ^{(2)}\left(\frac{i h}{\pi }+\frac{1}{2}\right)-\psi ^{(2)}\left(\frac{2 i h+2 i \eta+\pi }{2 \pi }\right)\right].
\end{equation}
with $h = \ln(\sqrt{c_0/c_{sf}})$ and $\psi^{(n)}$ being the $n$th-polygamma function. We can re-arrange the terms in Eq. (\ref{eq_LIV45}) to obtain,
%
\begin{eqnarray}
\label{eq28}
\langle(\Delta v^x)^2\rangle_{LIV} &=& \dfrac{2q^2B}{\pi m^2 c_0^4\eta_0^2}\left\lbrace\dfrac{1}{\pi^3}\left(\zeta(3)-\dfrac{\pi^4}{90}\right)+ \dfrac{\beta B^{-\frac{1}{2}}}{3}\left(\dfrac{c_0}{c_{sf}}\right)^{\frac{1}{2}} Re\left[F_k(\eta)\right]\right\rbrace, 
\end{eqnarray}
or, in terms of Eq. (\ref{caso ligado}), 
\begin{eqnarray}
\label{eq288}
\langle(\Delta v^x)^2\rangle_{LIV} &=& \langle(\Delta v^x)^2\rangle +  \frac{2q^2\beta B^{1/2}}{3\pi m^2 c_0^4\eta_0^2}\left(\frac{c_0}{c_{sf}}\right)^{1/2} Re\left[F_k(\eta)\right]
\end{eqnarray}
where the first term in the right-hand side is Eq. (\ref{caso ligado}) and the second term is the up to the first order corrections in $\beta$ that comes from the LIV model represented by Lagrangian (\ref{lagrangiana}). So the relevant correction contribution comes strictly from the theory represented by the Lagrangian (\ref{lagrangiana}), making a boost in the velocity dispersion.

Now let us plot Eq. (\ref{eq28}). Such a procedure can be done by defining a new function, $G(\eta)$, where
\begin{eqnarray}\label{eq:Gbound}
G(\eta) = \frac{\pi m^2c_0^4\eta_0^2}{2q^2B}\langle(\Delta v^x)^2\rangle_{LIV} &=& \frac{1}{\pi^3}\left(\zeta(3) - \frac{\pi^4}{90}\right) \\ \nonumber &+& \beta\left[\frac{3}{\pi^3}\left(\zeta(3) - \frac{\pi^4}{90}\right) + \frac{B^{-1/2}}{3}\left(\frac{c_0}{c_{sf}}\right)^{1/2}F_k(\eta)\right].
\end{eqnarray} 

\begin{figure}
\centering	
 \includegraphics[width=0.8\columnwidth]{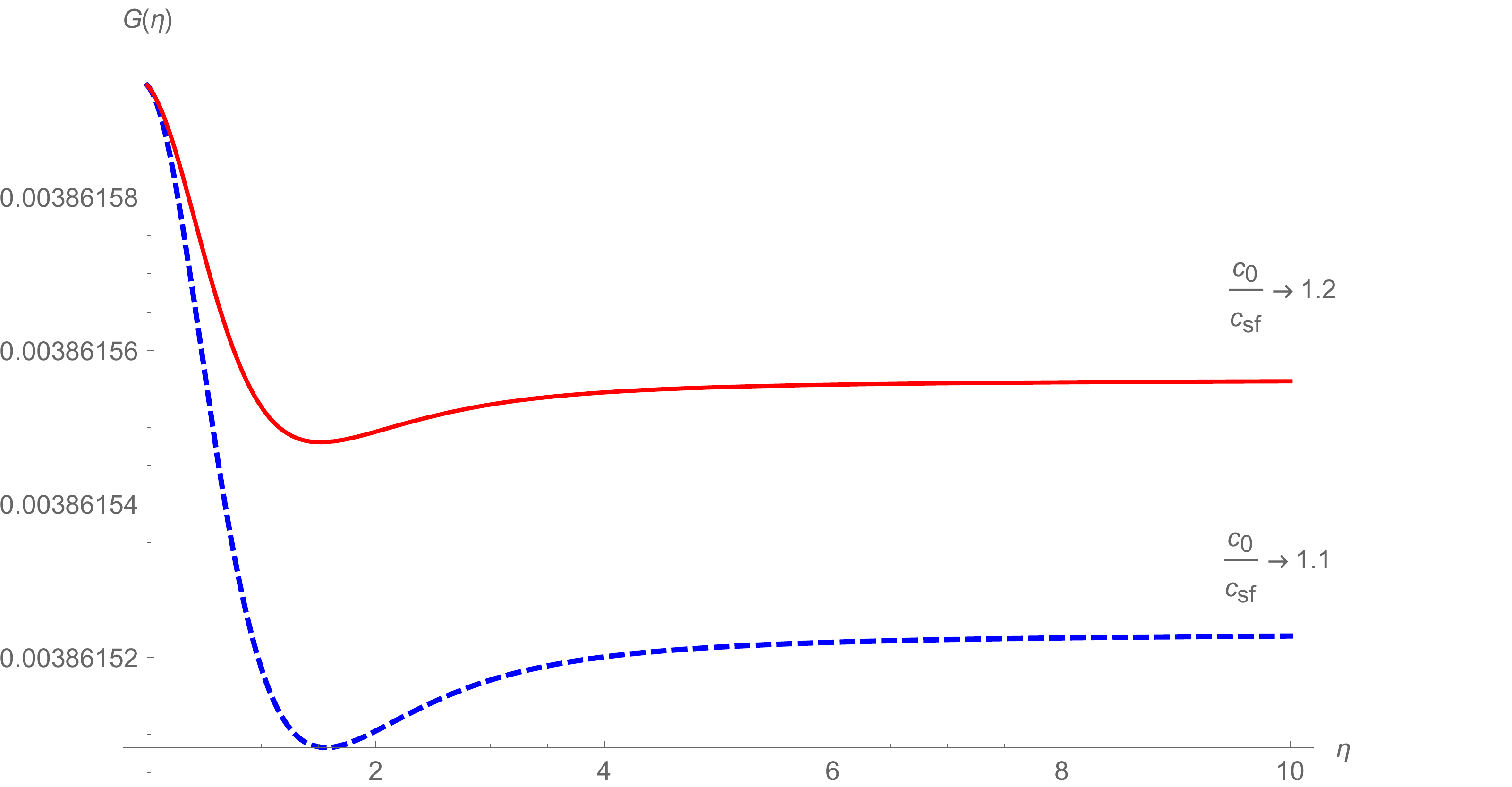} 
\caption{Dispersion for bounded particles for different values for the sound speed and $\beta=3.6\times 10^{-8}$. The dispersion decreases when  LIV is considered.   }\label{fig:bound}
 \end{figure}

In Fig. \ref{fig:bound} we plot the function $G(\eta)$ defined in Eq. (\ref{eq:Gbound}), which is directly related to $\langle(\Delta v^x)^2\rangle_{LIV}$. Here we choose $\beta = 3.6\times 10^{-8}$ as presented in Ref. \cite{mtn08}. Moreover, we have the constraint 
that $c_{sf} < c_0$, in order to the expansion occur. Given some numerical values to the sound speed we are able to evaluate the dimensionless quantities $B$ and $h$, both present in Eq. (\ref{eq:Gbound}). Thus, we have for the dashed curve $c_0/c_{sf} = 1.1$, $ B = 0.00227$ and $h = 0.0467$, whereas for the solid curve $c_0/c_{sf} = 1.2$, $ B = 0.00833$ and $h = 0.0911$. 

Note that, comparing the result found in Sect. \ref{Sect:review}, where LI is preserved, the  LIV decreases the value of the velocity dispersion when the particles are bounded for some external force. Decreasing, then, the uncertainty about this quantity. For large values of $\eta$ we see that $G (\eta)$ goes to a constant. In fact, as our time variable is found in Eq. (\ref{eq24}), taking the limit where $\eta \rightarrow \infty$, we have a constant value given by the polygamma function.

\subsection{Free particles}

In this section we will study the free particles case ($f^i_{ext} = 0$). The equation of motion (\ref{eq:motionLIV}) for a single particle can be expressed in the following way
\begin{equation}
\frac{1}{a^2}\dfrac{d}{dt}(a^2u^i)= \frac{q}{m}[a^{-2}(t)+\beta]\partial_i\phi_1,
\end{equation}
when $\beta = 0$,   we recover an equation valid for a flat or conformally flat spacetime with LI preserved.

The corresponding velocity-velocity correlation function is now
\begin{eqnarray}
\langle u^i(t_1, r_1)u^i(t_2, r_2)\rangle_{LIV} = \frac{q^2}{m^2a_f^4}\int\int dt_1dt_2a^2(t_1)a^2(t_2)\left\{\left[a^{-2}(t_1) + \beta\right]\left[a^{-2}(t_2) + \beta\right]\right\} \\ \nonumber \times \partial_{i_1}\partial_{i_2} \langle \phi_1(t_1, r_1)\phi_1(t_2, r_2)\rangle_{FRW}
\end{eqnarray}

Taking the following transformation in time, $dt = ad\eta$, we obtain the expression, 
\begin{eqnarray}
\langle u^i(t_1, r_1)u^i(t_2, r_2)\rangle_{LIV} &=&  \frac{q^2}{m^2a_f^4}\left[ \int_0^{\eta_f} d\eta_2\int_{0}^{\eta_f}d\eta_1  \partial_{1i}\partial_{2i}\langle\phi_1(\eta_1, r_1)\phi_1(\eta_2, r_2)\rangle_{M}\right.\nonumber\\
& + &\left. \beta  \int_0^{\eta_f} d\eta_2\int_{-\infty}^{\infty} d\eta_1 a^2(\eta_1) \partial_{1i}\partial_{2i}\langle\phi_1(\eta_1, r_1)\phi_1(\eta_2, r_2)\rangle_{M}\right.\nonumber\\
& + & \left. \beta  \int_{0}^{\eta_f} d\eta_1\int_{-\infty}^{\infty} d\eta_2 a^2(\eta_2) \partial_{1i}\partial_{2i}\langle\phi_1(\eta_1, r_1)\phi_1(\eta_2, r_2)\rangle_{M}\right. \nonumber\\
& + &\left. \beta^{2}\int_{-\infty}^{\infty} d\eta_2a^{2}(\eta_2)\int_{-\infty}^{\infty} d\eta_1a^{2}(\eta_1)\partial_{1i}\partial_{2i}\langle\phi_1(\eta_1, r_1)\phi_1(\eta_2, r_2)\rangle_{M}\right].
\end{eqnarray} 

Note that, the first term in the right hand side has no scale factor in the integrand. So a renormalization procedure is needed in this term. Thus, the first integral does not contributes and we have,
\begin{eqnarray}
\label{eq66}
\langle u^i(t_1, r_1)u^i(t_2, r_2)\rangle_{LIV} &=& \frac{q^2}{m^2a_f^4} \left[ \beta  \int_{0}^{\eta_f}d\eta_2\int_{-\infty}^\infty d\eta_1 a^2(\eta_1) \partial_{1i}\partial_{2i}\langle\phi_1(\eta_1, r_1)\phi_1(\eta_2, r_2)\rangle_{M}\right.\nonumber\\
& + & \left. \beta \int_{0}^{\eta_f}d\eta_1\int_{-\infty}^{\infty}d\eta_2 a^2(\eta_2) \partial_{1i}\partial_{2i}\langle\phi_1(\eta_1, r_1)\phi_1(\eta_2, r_2)\rangle_{M}\right. \nonumber\\
& + &\left. \beta^{2}\int_{-\infty}^{\infty}d\eta_2a^{2}(\eta_2)\int_{-\infty}^{\infty}d\eta_1a^{2}(\eta_1)\partial_{1i}\partial_{2i}\langle\phi_1(\eta_1, r_1)\phi_1(\eta_2, r_2)\rangle_{M}\right].
\end{eqnarray} 

Now putting Eq. (\ref{correlation}) into Eq. (\ref{eq66}) choosing $i = x$, the result for $y$ and $z$ is the same, and writing in terms of the proper velocity ($v^i$), where $v^i = a_fu^i$, we obtain,
\begin{eqnarray}
\label{eq66a}
\langle v^i(t_1, r_1)v^i(t_2, r_2)\rangle_{LIV} &=& \dfrac{q^2\beta }{2\pi^2m^2a_f^2}\int_{-\infty}^\infty d\eta_2 a^{2}(\eta_2)\int_{-\infty}^\infty d\eta_1 a^{2}(\eta_1)\left\{f_2(\eta,r)+4\Delta x^2f_3(\eta, r)\right\}+ \nonumber\\
&+& \dfrac{q^2 \beta}{2\pi^2m^2a_f^2}\int_{0}^{\eta_f} d\eta_2\int_{-\infty}^\infty d\eta_1 a^{2}(\eta_1)\left\{f_2(\eta,r)+4\Delta x^2f_3(\eta, r)\right\}+ \nonumber\\
&+& \dfrac{q^2 \beta^2}{2\pi^2m^2a_f^2}\int_{0}^{\eta_f} d\eta_1\int_{-\infty}^\infty d\eta_2 a^{2}(\eta_2)\left\{f_2(\eta,r)+4\Delta x^2f_3(\eta, r)\right\}. 
\end{eqnarray}
 with $f_n$ given by Eq. (\ref{eq:fn}).

Making use of Eq. (\ref{tanh}) and taking the coincidence limit we are able to make first an integration by parts and then apply the same contour integration of the previous sections to obtain,
\begin{equation}
\label{eq22a}
\langle(\Delta v^x)^2\rangle_{LIV} = \dfrac{2 q^2 \beta C}{3 \pi m^2 c_0^4 \eta_0^2}Re\left[F_k(\eta)\right]+\dfrac{2 q^2 \beta^2 B}{\pi^4 m^2 c_0^4 \eta_0^2}\zeta(3),
\end{equation}
with
\begin{equation}
\label{eq18a}
F_k(\eta)= \frac{14\zeta(3) + \psi^{(2)}\left(\frac{i\eta}{\pi} + \frac{1}{2}\right)}{2\pi^3}
\end{equation}
and $C = \frac{1}{2}\left(1 - \frac{c_{sf}}{c_0}\right).$ 
Again, the relevant contribution is the first-order correction in $\beta$ that comes from Lagrangian (\ref{lagrangiana}). This contribution produces a boost in the velocity dispersion that would vanish if $\beta = 0$.

Note that  a non-null dispersion was obtained which is different from the result found in Sect. \ref{Sect:review}. This is perhaps the most important result of the present manuscript, since this model shows a fundamental effect coming from the LIV theory. 

Now let us plot Eq. (\ref{eq22a}). Proceeding this way, define a new function, $G'(\eta)$, where
\begin{eqnarray}\label{eq:Gfree}
G'(\eta) = \frac{3\pi m^2c_0^4\eta_0^2}{2q^2}\langle(\Delta v^x)^2\rangle_{LIV} = C\beta F_k(\eta)
\end{eqnarray} 
\begin{figure}[htbp]
\centering	
 \includegraphics[width=0.8\columnwidth]{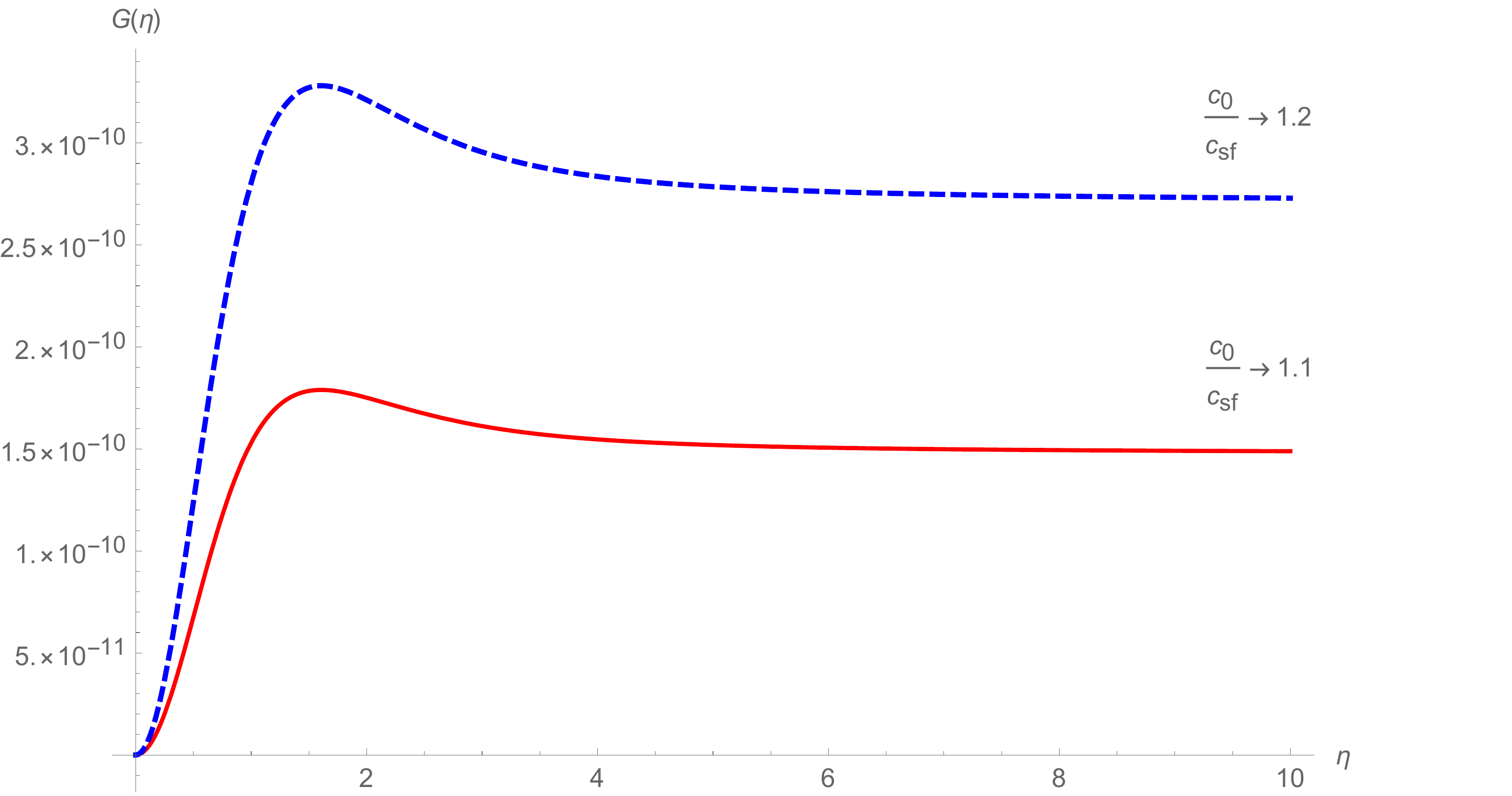}
 \caption{Dispersion for free particles for different values for the sound speed and $\beta=3.6\times 10^{-8}$. The dispersion increases when LIV is considered.}\label{Fig:free}
\end{figure}

The solid and dashed curves was plotted with $C = 0.04545$ and $C = 0.08333$, respectively, 
with the speed ratio $c_0/c_{sf} = 1.1$ and $1.2$. It was used $\beta = 3.6 \times 10^{-8}$ in both cases.
Note that this result is consistent with the free case described in Sect. \ref{Sect:review} only  when $\eta = 0$, since a null dispersion is found.
However, as the time increases, the dispersion is no longer zero, which disagrees with Sect. \ref{Sect:review}. This result is a direct consequence of the LIV. Thus, in the LIV scenario particles that follows their geodesics (free particles) can posses a stochastic motion due to quantum fluctuations. So LIV holds a fundamental influence in the non-trivial quantum effect described in this paper.

\section{Summary and discussion}\label{Sect:conclu}

In the present paper, we have considered the stochastic motion of scalar test particles coupled to a massless fluctuating scalar field in an acoustic spatially flat FRW  geometry. In this context, in Sect. \ref{Sect:review}, we reviewed the main results present in Ref. \cite{Bessa2017},  where to model this geometry, it was considered a fluid (BEC) where the LI was preserved. In this context, we studied stochastic motion in two distinct situations: in the first one, we have considered an external classical and non-fluctuating force acting on the particles in this geometry. This force prevented them from feel locally the expansion. We named them as bound particles. In the second situation, no external force was acting on the particles, in this case, they are free and follow their geodesics, and therefore they were named free particles. In particular, for the case studied in this section, the stochastic motion exists only for bound particles since for the free particles the dispersion vanishes ($\langle\Delta v^2\rangle = 0$) \cite{Bessa2017,Bessa2009}. 

In Sect. \ref{Sec:LIV},  following the same steps of Ref. \cite{abp10}, we showed that it is also possible to find an acoustic spatially flat FRW geometry where the LI is broken, {\it {i.e.}}, in a LIV context. Then, in Sect. \ref{sect:corrections}, we studied the same problem described in Sect. \ref{Sect:review} using the metric found in Sect. \ref{Sec:LIV}.  Thus, comparing the results found in Sect. \ref{Sect:review} with the ones found in Sect. \ref{sect:corrections}, we observed that for bound particles the LIV corrections decrease their velocity dispersion and also the uncertainty on the motion. We can see this fact from Eq. (\ref{eq288}) and Fig. \ref{fig:bound}. The reader can note that the curve for the dispersion decreases in the limit of long times. Another point worth mentioning is the dispersion relation between the sound speed in the beginning  ($c_0$) and the end of the expansion ($c_{sf}$). A change in the ratio ($c_0/c_{sf}$) causes a shift that might increase or decrease the dispersion. Nonetheless, the most important result in this paper is for the free particles in the LIV framework. This case is more realistic than the bound particles, and a null dispersion was found when the LI is taken into account. However, as can be seen, in Eq. (\ref{eq22a}) and in Fig. \ref{Fig:free}, when LIV is considered a non-null dispersion is found. We interpret this as a direct consequence of the LIV, and the dependence with the initial and final sound speeds is also illustrated in Fig. \ref{Fig:free}. 

Another important point related to our model appears in the $\beta$ corrections presented in Eqs. (\ref{eq288}) and (\ref{eq22a}). This term comes from Lagrangian (\ref{lagrangiana}), and it appears directly in the particle's equation of motion. Nevertheless, the acoustic metric plays a fundamental role in our results, not only because it is derived directly from Eq. (\ref{lagrangiana}) but also because in dispersion calculations the scale factor that appears in the integrands allows us to obtain finite and non-zero results.

As expected from a theory with LIV, the corrections found in the velocity dispersion of the particles are very small, and consequently, their detection is not an easy task. However, our model shows that the LIV plays a fundamental role related to the robustness of the motion of the particles, as in the case of the free particles, whereas the non-zero result obtained disagrees with the one previously found in the literature, where the Lorentz invariance is  preserved. Since experimental tests for quantum gravity effects are hard to find, the present model denotes a mean where such effects could be explored. Besides, the quantum stochastic motion is a subtle and non-trivial quantum scenario effect that is, per se, a topic of interest.

        \acknowledgments
     
				We would like to thank CNPq, CAPES and CNPq/PRONEX/FAPESQ-PB (Grants no. 165/2018 and 015/2019),  for partial financial support. MAA, FAB, EP and JRLS acknowledge support from CNPq (Grants no. 306962/2018-7 and  433980/2018-4, 312104/2018-9, 304852/2017-1, 420479/2018-0). The authors would like to thank J. P. Spinelly and F. G. Costa for helpful comments.  
         

\end{document}